\documentclass[10pt,twocolumn,aps,pra,showpacs,superscriptaddress,floatfix,nofootinbib]{revtex4-1}

\usepackage{graphicx}
\usepackage{amssymb}
\usepackage{amsmath}
\usepackage{color}
\usepackage{psfrag}
\usepackage{epsfig}
\usepackage{bbm}
\usepackage{bm}
\usepackage{hyperref}
\usepackage[normalem]{ulem}
\usepackage{amsthm}
\usepackage{epstopdf}

\newcommand{\beq}{\begin{eqnarray}}
\newcommand{\eeq}{\end{eqnarray}}

\definecolor{nblue}{rgb}{0.2,0.2,0.7}
\definecolor{ngreen}{rgb}{0.2,0.6,0.2}
\definecolor{nred}{rgb}{0.8,0.2,0.2}
\definecolor{nblack}{rgb}{0,0,0}
\definecolor{hotmagenta}{rgb}{1.0, 0.11, 0.81}

\DeclareMathOperator{\tr}{tr}

\theoremstyle{definition}

\begin{document}
\title{Hierarchy in temporal quantum correlations}

\author{Huan-Yu Ku}
\affiliation{Department of Physics, National Cheng Kung University, 701 Tainan, Taiwan}
\affiliation{Theoretical Quantum Physics Laboratory, RIKEN Cluster for Pioneering Research, Wako-shi, Saitama 351-0198, Japan}

\author{Shin-Liang Chen}
\affiliation{Department of Physics, National Cheng Kung University, 701 Tainan, Taiwan}
\affiliation{Max-Planck-Institut f{\"u}r Quantenoptik, Hans-Kopfermann-Stra{\ss}e 1, 85748 Garching, Germany}

\author{Neill Lambert}
\affiliation{Theoretical Quantum Physics Laboratory, RIKEN Cluster for Pioneering Research, Wako-shi, Saitama 351-0198, Japan}


\author{Yueh-Nan Chen}
\email{yuehnan@mail.ncku.edu.tw}
\affiliation{Department of Physics, National Cheng Kung University, 701 Tainan, Taiwan}
\affiliation{Theoretical Quantum Physics Laboratory, RIKEN Cluster for Pioneering Research, Wako-shi, Saitama 351-0198, Japan}
\affiliation{Physics Division, National Center for Theoretical Sciences, 300 Hsinchu, Taiwan}

\author{Franco Nori}
\affiliation{Theoretical Quantum Physics Laboratory, RIKEN Cluster for Pioneering Research, Wako-shi, Saitama 351-0198, Japan}
\affiliation{Department of Physics, The University of Michigan, Ann Arbor, 48109-1040 Michigan, USA}

\date{\today}

\begin{abstract}
Einstein-Podolsky-Rosen (EPR) steering is an intermediate quantum correlation that lies in between entanglement and Bell non-locality. Its temporal analog, temporal steering, has recently been shown to have applications in quantum information and open quantum systems. Here, we show that there exists a hierarchy among the three temporal quantum correlations: temporal inseparability, temporal steering, and macrorealism. Given that the temporal inseparability can be used to define a measure of quantum causality, similarly the quantification of temporal steering can be viewed as a weaker measure of direct cause and can be used to distinguish between direct cause and common cause in a quantum network.
\end{abstract}

\maketitle

\section{Introduction}
The concept of quantum steering was first articulated by Shr\"odinger~\cite{Schrodinger35} in response to the apparently non-local phenomenon of quantum correlations questioned by Einstein, Podolsky, and Rosen (EPR)~\cite{EPR35}. Thanks to the celebrated inequality proposed by Bell~\cite{Bell64}, a great deal of theoretical and experimental investigation has been focused on quantum non-locality in the past few decades. Empowered by practical quantum information task requirements, spatial EPR steering was recently able to be studied in a more quantitative way~\cite{Wiseman07,Jones07,Cavalcanti09,Smith12,Bernhard12}. Together with the concepts of Bell nonlocality and entanglement, EPR steering forms a hierarchy, and as such acts as an intermediate quantum correlation that lies in between the others~\cite{Wiseman07,Jones07,Cavalcanti09}, i.e., EPR steering is, in general, weaker than Bell nonlocality but stronger than quantum entanglement. Research on EPR steering in the past few years has seen the development of several interesting new avenues of study~\cite{Rodrigo15,Skrzypczyk14,Piani15,Cavalcanti17,Cavalcanti16,Uola14,Quintino14,Uola15,YNChen14,SLChen16_2,Bartkiewicz16,HYKu16,Li15,Xiong17,Costa17}. In addition to these theoretical developments, EPR steering has also been observed experimentally~\cite{Guerreiro16,Wittmann12,Bennet12}.

The notion of causality, cause and effect, is an intuitive concept. In quantum mechanics, however, applying the concept of causality is not always that straightforward. For example, quantum mechanics allows the superposition principle to be applied to causal relations, such that indefinite casual order may occur with proper design~\cite{Chiribella13,Oreshkov12}. A measurement of a superposition of causal orders has been demonstrated very recently~\cite{Rubinoe17}. Another driving force for the research on quantum causality~\cite{Brukner14} comes from Bell's theorem, and its generalizations, that can be analyzed with a causal approach~\cite{Tobias12,Wood15,Chaves14,Chaves15}. Potential applications of quantum casual relations in quantum information tasks have also been proposed~\cite{Hardy2009,Chiribella12,Mateus14,Procopio15}.

In contrast to creating an indefinite causal order, some other experimental works related to quantum causal relations have also attracted attention, e.g. distinguishing different causal structures (common cause and direct cause)~\cite{Ried15} and defining a measure of quantum causal effects (direct cause)~\cite{Fitzsimons15}.

Our goal in this work is to relate temporal steering to the notion of a quantum causal effect. To do so we first show that there also exists a hierarchy among the three temporal quantum correlations (temporal inseparability, temporal steering, and nonmacrorealism), which are provided when the condition of no-signalling in time (NSIT)~\cite{Brukner1, Cheming,Brukner2} is obeyed. When NSIT in temporal steering is violated, non-vanishing temporal steering may occur under a dephasing process, which we prove to be the same as the  distinguishability between two purely classical assemblages. Given that the temporal inseparability can be used to define a measure of quantum causal effects, we conclude that temporal steering can be viewed as a weaker measure of quantum causal effect and can be used to distinguish between direct cause and common cause in a quantum network.

\section{Macrorealism} Consider a system that evolves with time, and on which one can measure a physical quantity $Q$ at time $t_1$, $t_2$, or $t_3$ to obtain the corresponding values $Q(t_1)$, $Q(t_2)$, and $Q(t_3)$, respectively. In 1985, Leggett and Garg (LG)\cite{Leggett1985,Emary2014} proposed an inequality: 
\begin{equation}
K\equiv C'_{12} + C'_{23} - C'_{13} \leq 1,
\end{equation} where $C_{ij}\equiv \langle Q(t_i) Q(t_j)\rangle $ is the expectation value of the measurement outcomes at time $t_i$ and $t_j$~\cite{emaryPRL,lambertPRA}. This inequality holds if the dynamics of the system is classical, in the realism sense, and the measurements are non-invasive. Violation of the inequality shows the incompatibility between quantum mechanics and macrorealism.

\begin{figure}[tbp]
\includegraphics[width=1\columnwidth]{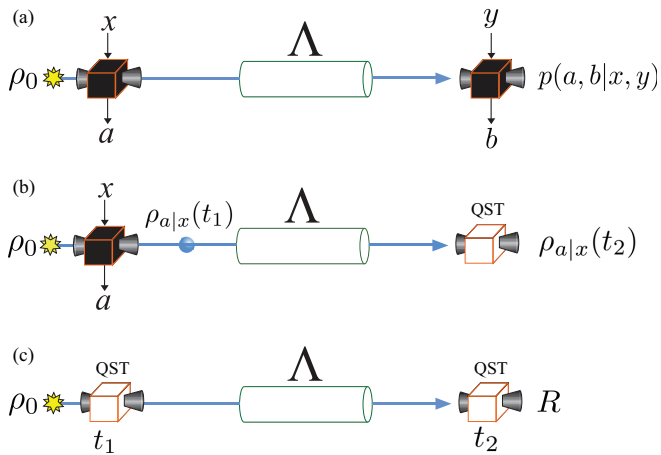}
\caption{Beginning with quantum state $\rho_0$, one can construct three different temporal quantum correlations: (a) the temporal correlations scenario associated with nonmacrorealism, (b) the temporal steering scenario, and (c) constructing the pseudo density matrix of a single system.
Here, the $x$ and $y$ denote two classical inputs at time $t_1$ and $t_2$ with classical outcomes $a$ and $b$, respectively.  The quantum channel is denoted by $\Lambda$. Quantum state tomography is denoted as QST.}
\label{fig_TCHSH}\centering %
\end{figure}

One can consider a more general scenario to investigate temporal correlations. For instance, there can be two or more quantities being measured at each moment of time. For simplicity, we consider the scenario with two times $t_1=0$ and $t_2=t$, at which the quantity $x\in\{x\}_{x=1}^{n_x}$ and the quantity $y\in\{y\}_{y=1}^{n_y}$ are measured respectively during each round of the experiment. Accordingly, one obtains the outcome $a\in\{a\}_{a=1}^{n_a}$ and the outcome $b\in\{b\}_{b=1}^{n_b}$ (see Fig.~\ref{fig_TCHSH}). After many rounds of the experiment, one can obtain a set of probability distributions $\{p(a,b|x,y)\}_{a,b,x,y}$. Then, a macrorealistic (MS) theory restricts the probability distributions to be of the following form:
\begin{equation}
p(a,b|x,y)\stackrel{\text{MS}}{=}\sum_\lambda p(a|x,\lambda)~p(b|y,\lambda)~p(\lambda)~~~\forall a,b,x,y.
\label{Eq_MS}
\end{equation}
The physical interpretation of the above equation is the following: The probability distribution $p(a,b|x,y)$ between time $t_1$ and $t_2$ does not depend on the history of the experiment. Therefore, there exist hidden parameters $\lambda$, which can be deterministic or stochastic~\cite{Clemente15}, defining all physical properties and forming the probability distributions $p(a|x,\lambda)$ and $p(b|y,\lambda)$.


In quantum theory, a measurement outcome is typically not pre-determined due to intrinsic uncertainty. The probability distributions follow Born's rule: 
\begin{equation}
p(a,b|x,y)\stackrel{\mathcal{Q}}{=}\text{tr}\left[E_{b|y}\cdot \mathcal{E} (\sqrt{E_{a|x}}~\rho_0\sqrt{E_{a|x}})\right] ~\forall a,b,x,y, 
\end{equation} where $\rho_0$ is the initially prepared quantum state, $\{E_{a|x}\}_a$ denotes the positive-operator valued measurement (POVM), $E_{a|x}\geq 0$, $\sum_{a}E_{a|x}=\openone$ of each $x$, similarly $\{E_{b|y}\}_b$ is POVM of each $y$, and $\mathcal{E}$ describes the dynamics of the system from $t_1=0$ to $t_2=t$. In the following, the sets of probability distributions which do not admit Eq.~(\ref{Eq_MS}) will be called \emph{nonmacrorealistic}.

Similar to the spatial case, one can also write down the so-called \emph{temporal Bell inequalities}~\cite{Fritz10} to be a set of constraints for the macroscopical probability distributions. For instance, setting $n_x=n_y=n_a=n_b=2$ and shifting $a,b \in \{1,2\}$ to $a,b\in\{\pm 1\}$, the temporal Clauser-Horne-Shimony-Holt (CHSH) kernel is written as \begin{equation}
\langle B\rangle_{\text{T-CHSH}} \equiv C_{xy} +C_{x'y} + C_{xy'} - C_{x'y'},
\end{equation} where 
\begin{equation}
C_{xy}\equiv p(a=b|x,y)-p(a\neq b|x,y)
\end{equation}
 is the expectation value of $a\cdot b$. For a qubit system, $\langle B\rangle_{\text{T-CHSH}}$ is upper bounded by $2$ and $2\sqrt{2}$ for the MS model and quantum mechanics, respectively.

To give a proper quantification of the degree of nonmacrorealistic dynamics, we follow the techniques used for standard Bell inequalities, i.e., optimizing all possible combinations of the measurement settings which give the maximal quantum violation of $\langle B\rangle_{\text{T-CHSH}}$:
\begin{equation}
\langle B\rangle_{\text{T-CHSH}}^\text{max} =  \max_{x,x',y,y'} \Big\{0, ~\frac{\langle B\rangle_{\text{T-CHSH}}-2}{2\sqrt{2}-2} \Big\}.
\end{equation}

\section{Temporal steering} Now, consider that one can perform  quantum state tomography (QST) to obtain the quantum state at time $t_2=t$ instead of obtaining the probability distributions. After many rounds of the experiment, one can obtain a set of quantum states $\{\hat{\sigma}_{a|x}(t)\}$ corresponding to those states found after the measurement event $a|x$ at time $t_1=0$. It is rather convenient to define the so-called \emph{temporal assemblage} as a set of subnormalized state $\{\rho_{a|x}(t)\equiv p(a|x)~\hat{\sigma}_{a|x}(t)\}$. Through this, a temporal assemblage contains the information on both $p(a|x)=\tr[\rho_{a|x}(t)]$ and $\hat{\sigma}_{a|x}(t) = \rho_{a|x}(t)/\tr[\rho_{a|x}(t)]$.
If one believes the measurement at time $t_1=0$ is \emph{noninvasive}, i.e., knowing the outcome $a$ in \emph{prior}, without disturbing the system and its subsequent dynamics, then the observed temporal assemblage should satisfy the hidden-state model~\cite{YNChen14,SLChen16_2,Thesis_Chen}
\begin{equation}
\rho_{a|x}(t)\stackrel{\emph{\rm{noninvasive}}}{=}\sum_{\lambda}p(\lambda)p(a|x,\lambda)\sigma_{\lambda}~~~\forall a,x.
\label{Eq_HSM}
\end{equation}
The physical interpretation of the temporal hidden-state model is the following: During each experimental round, there exists an ontic state $\lambda$, which predetermines the outcome $a$ when performing the measurement $x$ at $t_1=0$, as well as pre-determining the quantum state $\sigma_\lambda$ at time $t_2=t$.

A temporal assemblage which admits a quantum mechanical model can be written as: 
\begin{equation}
\rho_{a|x}(t)\stackrel{\mathcal{Q}}{=}\mathcal{E} (\sqrt{E_{a|x}}\rho_0\sqrt{E_{a|x}}).
\end{equation}
Given a temporal assemblage, one can know if it admits the hidden-state model Eq~\eqref{Eq_HSM} by the feasibility problem of 
\begin{equation}
\{ \text{find}~\sigma_\lambda~|~\rho_{a|x}(t)=\sum_{\lambda}p(\lambda)p(a|x,\lambda)\sigma_{\lambda}\}.
\end{equation}

We refer to those assemblages, which do not admit the hidden-state model, as \emph{temporal steerable}, and the degree of the temporal steerability is quantified by the measure of \emph{temporal steerable weight}~\cite{SLChen16_2} and \emph{temporal steering robustness} (TSR)~\cite{HYKu16}. In the following, we will use TSR to quantify the degree of temporal steerability for a given temporal assemblage: 
\begin{equation}
\begin{aligned}
&\text{TSR}=\text{min}~\alpha~~ \text{subject to}\\&\left\{\frac{1}{1+\alpha}\rho_{a|x}(t)+\frac{\alpha}{1+\alpha}\tau_{a|x}(t)=\sum_{\lambda}p(\lambda)p(a|x,\lambda)\sigma_{\lambda}\right\}_{a,x}
\end{aligned}
\end{equation}
 where $\tau_{a|x}(t)$ is a valid noisy temporal assemblage. This can be formulated as a semidefinite programming problem~(SDP)~\cite{SDP,Pusey13,Skrzypczyk14,SLChen17,SLChen16} as follows:
\begin{equation}
\begin{aligned}
\text{TSR}&=~\text{min}\left(\text{tr}\sum_{\lambda}\sigma_{\lambda} -1  \right),~~\text{with}~~\sigma_{\lambda}\geq 0~~ \forall~\lambda\\
&\text{subject~~to}~~\sum_{\lambda}p(a|x,\lambda)\sigma_{\lambda}-\rho_{a|x}(t)\geq 0~~\forall~ a,x.
\end{aligned}
\end{equation}

\section{Pseudo density matrix and temporal inseparability} To complete the picture of a hierarchy of correlations, we give a brief introduction to the so-called \emph{pseudo density matrix} introduced by Fitzsimons \emph{et al.}~\cite{Fitzsimons15}. A pseudo density matrix is a way to define the state of one (or more) system between two (or more) moments of time. By definition, the pseudo density matrix $R$ of a qubit passing through a quantum channel is obtained by performing the QST before and after the evolution (see Fig.~\ref{fig_TCHSH}). Therefore, the pseudo density matrix is expressed as:
\begin{equation}
R = \frac{1}{4}\sum_{i,j=0}^3 C_{ij} \cdot \sigma_i\otimes\sigma_j,
\end{equation}
where $\{\sigma_i\}_{i=0,1,2,3} = \{\openone, \hat{X}, \hat{Y}, \hat{Z}\}$ is the set composed of the identity operator and the Pauli matrices.
Here, $C_{ij}=\text{tr}(R\cdot \sigma_i\otimes\sigma_j)$ are the expectation values of the result of these quantum measurements.
A pseudo density matrix is hermitian and normalized, but not necessarily positive-semidefinite. In general, a pseudo density matrix can also describe the state between two systems at different time. One can see that $R$ becomes a standard density matrix, which is positive-semidefinite, when the time-separation $t_2-t_1=0$. Therefore, the relation between two measurement events is called \emph{space-like} correlated when $R$ is positive-semidefinite.

Conversely, if $R$ is not positive-semidefinite, it is definitely not constructed from a standard spatially separated system. In this case, the relation between two measurement events is called \emph{time-like} correlated. In Ref.~\cite{Fitzsimons15}, the authors proposed a measure, called the $f$\emph{-function}, to quantify the degree of such a temporal relation:
\begin{equation}
f = \sum_i |\mu_i|,
\end{equation}
which is the summation over all the negative eigenvalues $\{\mu_i\}$ of a given $R$. In the rest of the discussions, all the pseudo density matrices are obtained by considering a single qubit at different times.
Due to the mathematical similarity to a separable quantum state, in the following we will refer to the situation $f\neq 0$ as \emph{temporally inseparable}. It is worth noting that the ``separability" here does not denote the separability with respect to two spatially separable systems, but indicates the pseudo density matrix can be written in the separable form $R = \sum_\lambda p(\lambda) \omega_\lambda^\text{A} \otimes\theta_\lambda^{\bar{\text{A}}}$, where $p(\lambda)$ is the probability distribution, $\omega_\lambda^\text{A}$ and $\theta_\lambda^{\bar{\text{A}}}$ are some valid quantum states acting on Hilbert spaces $\mathcal{H}_\text{A}$ at $t_1=0$ and $\mathcal{H}_{\bar{\text{A}}}$ at $t_2=t$, respectively~\cite{footnote}. Note that, in general, a temporal separable model implies $f=0$, but not vice versa.

\section{A hierarchy of temporal quantum correlations} Now we show a hierarchical relation between three temporal relations: nonmacrorealism, temporal steerability, and temporal inseparability. To this end, we show one can obtain the temporal assemblage $\{\rho_{a|x}(t)\}_{a,x}$ by performing a set of POVMs $\{E_{a|x}\}_{a,x}$ on the pseudo density matrix $R$, in which $\{E_{a|x}\}_{a,x}$ are the POVMs producing $\{\rho_{a|x}(t)\}_{a,x}$. More precisely, we show
\begin{equation}
\rho_{a|x}(t) = \text{tr}_\text{A}(E_{a|x}\otimes\openone\cdot R),
\label{Eq_TS_from_PDM}
\end{equation}
where 
\begin{equation}
\rho_{a|x}(t)=\mathcal{E}(\rho_{a|x}(0))=\mathcal{E}\left(\sqrt{E_{a|x}}~\rho_0\sqrt{E_{a|x}}\right)
\end{equation}
and $\rho_0=\openone/2$. The proof is given in Appendix A. Once Eq.~\eqref{Eq_TS_from_PDM} holds, the following formulation of an assemblage can be derived:
\begin{equation}
\begin{aligned}
\rho_{a|x}(t)
&= \tr_\text{A}(E_{a|x}\otimes\openone\cdot\sum_\lambda p(\lambda)\omega_\lambda^\text{A}\otimes\theta_\lambda^{\bar{\text{A}}})\\
&= \sum_\lambda p(\lambda)p_Q(a|x,\lambda)\theta_\lambda^{\bar{\text{A}}},
\end{aligned}
\end{equation}
where the set of probabilities $p_Q(a|x,\lambda):=\text{Tr}(E_{a|x}\omega_\lambda^\text{A})$ is constrained by the uncertainty relation~\cite{Maassen88}. In Eq. (7), we assume that the pseudo density matrix $R$ is temporally separable. Since the set of probability distributions $p(a|x,\lambda)$ in a hidden-state model Eq.~\eqref{Eq_HSM} is only constrained by the normalization property, a hidden-state model can reproduce an assemblage given by the above equation, but not vice versa. Therefore, we arrive at the hierarchical relation between temporal separability and temporal hidden-state model: a temporal assemblage constructed from a dynamical evolution admits a temporal hidden-state model if the corresponding pseudo-density matrix is temporally separable. Similarly,
\begin{equation}
\begin{aligned}
p(a,b|x,y)
&= \tr\left[E_{b|y}\cdot\rho_{a|x}(t)\right]\\
&= \tr\left[E_{b|y}\cdot\sum_\lambda p(\lambda)p(a|x,\lambda)\sigma_\lambda\right]\\
&= \sum_\lambda p(\lambda)p(a|x,\lambda)p_Q(b|y,\lambda)
\end{aligned}
\end{equation}
can be reproduced by macroscopic correlations [Eq.~\eqref{Eq_MS}], but not vice versa, i.e., there is a hierarchical relation between the temporal hidden-state model and macrorealism: a temporal correlation is macrorealistic if the corresponding temporal assemblage admits a temporal hidden-state model. The hierarchy can be described in a converse way: a nonmacrorealistic dynamics leads to a temporal steerable assemblage, and a dynamics which leads to a temporal steerable assemblage gives an inseparable pseudo density matrix.

In the following, we propose a proposition which will be used in the example.

\textbf{Proposition.} When the initial state of the qubit is prepared in the maximally mixed state, the pseudo density matrix constructed under the amplitude-damping, the phase-damping, and the depolarizing channels are temporal-separable if $f=0$, i.e.,
\begin{equation}
\begin{aligned}
f = 0 \Rightarrow ~~&\exists~~ \omega_\lambda^\text{A}, \theta_\lambda^{\bar{\text{A}}} ,~
\text{such that}~ R = \sum_\lambda p(\lambda) \omega_\lambda^\text{A} \otimes\theta_\lambda^{\bar{\text{A}}}.
\end{aligned}
\label{Eq_temporal_sep}
\end{equation}
The purpose of using the maximally mixed state as the initial condition is to produce assemblages which admit NSIT (cases which violate NSIT will be discussed later). The proof of the proposition is given in Appendix B.

\begin{figure}[tbp]
\includegraphics[width=1\columnwidth]{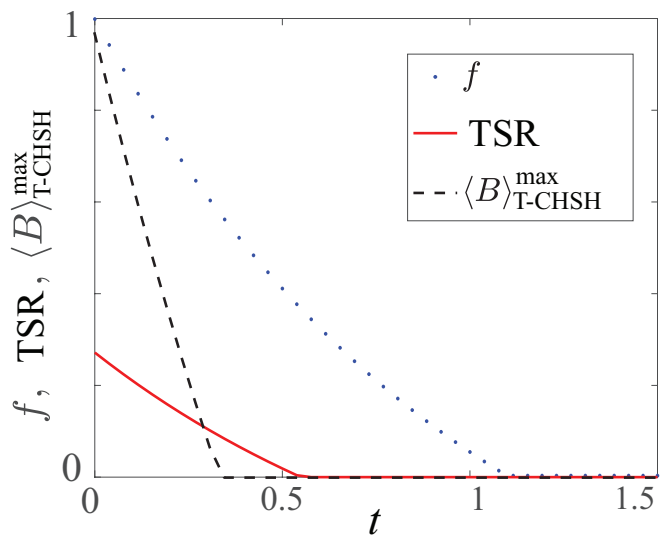}
\caption{The blue-dotted, red-solid, and black-dashed curves represent, respectively, the dynamics of the $f$, the TSR, and $\langle B\rangle_\text{T-CHSH}^{\text{max}}$ of a qubit undergoing the depolarizing channel~Eq.~\eqref{Eq_master_D}. We can see that the order of the three quantifiers (from the earliest to the latest vanishing time) is $\langle B\rangle_\text{T-CHSH}$, TSR, and $f$, demonstrating the hierarchy relation proposed in this work. Here, $t$ is in units of depolarizing rate $\gamma_\text{D}$.}
\label{fig_hierarchy}\centering %
\end{figure}

As a simple example, we consider a qubit experiencing a depolarizing channel, described by Eq.~\eqref{Eq_master_D}. In Fig.~\ref{fig_hierarchy}, we plot the dynamics of the $f$, the TSR, and $\langle B\rangle_{\text{T-CHSH}}^{\text{max}}$. We can see that the vanishing time, in which the corresponding classical model emerges, of each quantifier is different, demonstrating the hierarchical relation among the three temporal quantum relations.

\section{Classical steering} In the above discussions, the scenario we consider is under the condition of NSIT. That is, the obtained temporal assemblages $\{\rho_{a|x}(t)\}$ obey 
\begin{equation}
\sum_a \rho_{a|x}(t)=\sum_a\rho_{a|x'}(t)~~\forall x\neq x'.
\end{equation}
Given that temporal hidden-state model and temporal Bell inequalities assume non-invasive measurements, observing non-macrorealism and temporal steering while satisfying NSIT gives a stricter example of both properties \cite{halliwell1,halliwell2} (in that it rules out certain types of examples of false signatures of both effects due to classical clumsiness). In this section, we give a simple example of such a false-signature, which appears when the obtained temporal assemblages are not restricted to NSIT.

First, we show that, by the following explicit example, instead of performing measurements on an initial quantum state, one can prepare a temporal assemblage which leads to temporal steerability by just preparing a set of ``classical (subnormalized) states":
\begin{equation}
\rho_{a|x}(0)=\text{diag}[\alpha_{a|x},\beta_{a|x}],
\end{equation}
where $\alpha_{a|x}$ and $\beta_{a|x}$ are non-negative real numbers with $a$, $x\in \{1,2\}$. We refer to these states as ``classical" since all of them have just diagonal terms. Therefore, each state can be created by mixing, say, the spin of electrons in just one direction (e.g., $z$-direction). Such a temporal assemblage is steerable but trivial, and this is the reason that this scenario is not considered in the previous discussion, and ruled out by assuming NSIT.

In Appendix C, we show that if the measurement settings at time $t_1=0$ are set to be two, the asymptotic value of TSR (or temporal steerable weight) when time goes to infinity will be the same as the trace distance between the summation of the elements of the temporal assemblage in different measurement settings, i.e.,
\begin{equation}
\text{TSR}(\{\rho_{a|x}(t\rightarrow\infty)\}) = D\left(\sum_{a}\rho_{a|x}~,~\sum_{a}\rho_{a|x'}\right),
\label{Eq_TSR_traceD}
\end{equation}
where $D$ is the trace distance between two quantum states.
One notes that the trace distance in the classical case represents the distinguishability between two probability distributions. Equation~\eqref{Eq_TSR_traceD} means that the quantification of temporal steering arises from a classically ``clumsy" experiment if the condition of NSIT is violated.



\begin{figure}[tbp]
\includegraphics[width=1\columnwidth]{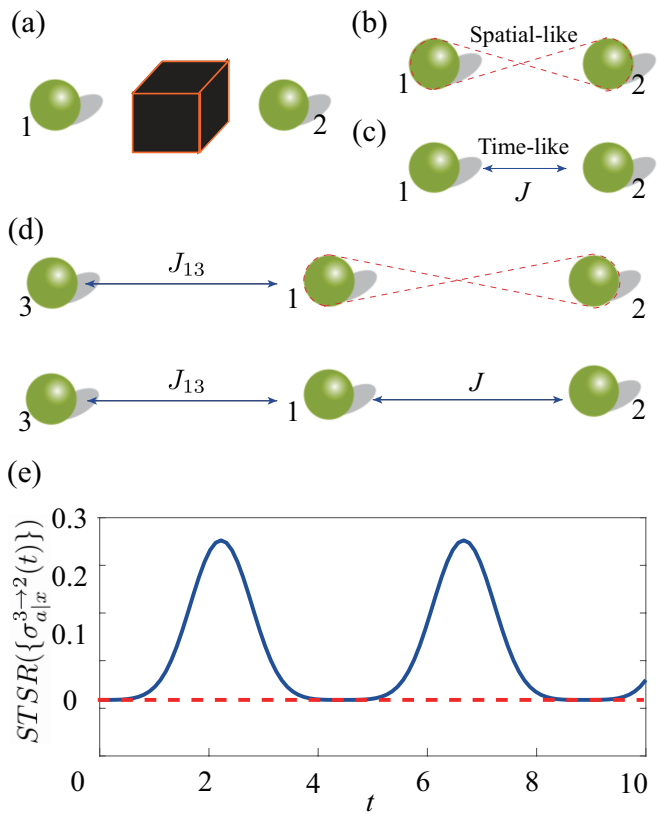}
\caption{Schematic diagram for the quantum causality schemes considered here. (a) The quantum correlations of two qubits arise from a ``black box''. The correlations may be due to (b) a common cause (space-like correlation) or (c) a direct cause (time-like correlation). (d) and (e): We propose to include an auxiliary qubit (qubit-3) coherently coupled to qubit-1. By examining the time-like steering of qubit-3 to qubit-2, one can infer whether the correlations are due to a common cause or a direct cause. This is because, if the correlations are from the common cause, there is no time-like steering of qubit-3 to qubit-2 (dashed line), while an oscillatory time-like steering (blue curve) exists if they are from a direct cause.}
\label{fig_three_qubits_combined}\centering %
\end{figure}

\section{Inferring causal structure with temporal steerability} Finally, motivated by Ref.~\cite{Fitzsimons15} proposing the $f$-function as a measure of quantum causal effect, which discriminates between spatial and temporal correlations, we propose that the degree of temporal steeribility can also be another measure of a quantum causal effect.

First of all, let us define the scenario of quantum causality discussed here. Consider two quantum systems that interact with each other through a black box as shown in Fig.~3(a). The correlations between the two systems may be due to  spatial correlations (common cause)  in Fig.~3(b) or temporal correlations (direct cause) in Fig.~3(c). The problem we would like to address is that how to discriminate between these two scenarios without knowing the mechanism of the black box.

To illustrate that temporal steering can discriminate between common and direct cause, we propose to include an auxiliary qubit (qubit-3) coherently coupled to qubit-1 as shown in Fig.~3(d). For illustrative purpose, we consider the following two scenarios. The first scenario is that qubit-1 and -2 initially share a maximally entangled state, while, for the second scenario, qubit-1 and -2 are coherently coupled with each other via the Hamiltonian \begin{equation}
H=\hbar J(\sigma_1^+\sigma_2^- + \sigma_1^-\sigma_2^+),
\end{equation}
where $J$ is the coupling strength and $\sigma_i^\pm$ are the raising and lowering operators of qubit-$i$.
To obtain the temporal assemblage of qubit-2 at $t_2=t$, three measurements in mutually unbiased bases of $\hat{X}$, $\hat{Y}$, and $\hat{Z}$ are performed on qubit-3 at time $t_1=0$. Actually, this is the so-called \emph{spatio-temporal steering} scenario~\cite{SLChen17}, which is a generalization of temporal steering. The TSR of qubit-2  $\text{TSR}_{3\rightarrow 2}(\{\rho_{a|x}^{2}(t)\})$ is plotted in Fig.~\ref{fig_three_qubits_combined}. We can see that in the case that two qubits share a common cause, $\text{TSR}_{3\rightarrow 2}(\{\rho_{a|x}^{2}(t)\})$ is always zero, while in the case that two qubits are connected by a direct cause, $\text{TSR}_{3\rightarrow 2}(\{\rho_{a|x}^{2}(t)\})$ oscillates with time. This simple example illustrates how, as one might expect, given the hierarchy of temporal correlation introduced earlier, that the temporal steerability can be used to distinguish between the direct and common causal effect in a quantum network.

\section{Concluding Remarks} It is worth to note that a hierarchy relation between temporal steerability and macrorealism is also considered in Ref.~\cite{Mal15}. However, in their work, neither the steerability witness nor the temporal CHSH inequality is optimized. The results of our work fill this gap. Open questions include: does the separable property of proposition Eq.~\eqref{Eq_temporal_sep} hold for any quantum channel? Can a temporal assemblage be obtained directly from a pseudo density matrix under the requirement of the violation of no-signaling in time? How will the hierarchical relation change if we consider another formulation of ``a state over time", e.g., the one in~\cite{Leifer13} or the one constructed by a discrete Wigner representation~\cite{Wootters87,Gross06} (see Ref.~\cite{Horsman16} for more comparisons between the three methods.)

\begin{acknowledgments}
The authors acknowledge useful discussions with Adam Miranowicz and Roope Uola. 
H.-Y. K. acknowledges the support of the Graduate Student Study Abroad Program (Grants No. MOST 107-2917-I-006-002). S.-L. C. would like to acknowledge the support from Postdoctoral Research Abroad Program (Grants No. MOST 107-2917-I-564 -007).
This work is supported partially by the National Center for Theoretical Sciences and Ministry of Science and Technology, Taiwan, Grants No. MOST 103-2112-M-006-017-MY4. 
F.N. is supported in part by the: MURI Center for Dynamic Magneto-Optics via the Air Force Office of Scientific Research (AFOSR) (FA9550-14-1-0040), Army Research Office (ARO) (Grant No. 73315PH), Asian Office of Aerospace Research and Development (AOARD) (Grant No. FA2386-18-1-4045), Japan Science and Technology Agency (JST) (the ImPACT program and CREST Grant No. JPMJCR1676), Japan Society for the Promotion of Science (JSPS) (JSPS-RFBR Grant No. 17-52-50023). N.L. and F.N. are supported by the Sir John Templeton Foundation and the RIKEN-AIST Challenge Research Fund.

\emph{Note added.---} Recently we became aware of \cite{Uola17}, which independently proved a hierarchy between temporal steerability and nonmacroscopicity.
\end{acknowledgments}

\appendix

\section{Obtaining a set of temporal correlations and a temporal assemblage from a pseudo density matrix}\label{App_hierarchy}
Now, we show that one can obtain the temporal assemblage $\{\rho_{a|x}(t)\}_{a,x}$ by performing a set of positive-operator valued measurements (POVMs) $\{E_{a|x}\}_{a,x}$ with $E_{a|x}\geq 0$, satisfying
$\sum_{a}E_{a|x}=\openone$, on the pseudo density matrix $R$, in which $\{E_{a|x}\}_{a,x}$ is the POVMs producing $\{\rho_{a|x}(t)\}_{a,x}$. More precisely, we show
\begin{equation}
\rho_{a|x}(t) = \text{tr}_\text{A}(E_{a|x}\otimes\openone~R)~~\forall a,x,
\label{App_Eq_TS_from_PDM}
\end{equation}
where \begin{equation}
\rho_{a|x}(t)=\mathcal{E}(\rho_{a|x}(0))=\mathcal{E}\left(\sqrt{E_{a|x}}\rho_0\sqrt{E_{a|x}}\right)
\end{equation} and $\rho_0=\openone/2$.

\emph{Proof.---} Without the loss of generality, we assume $\{E_{a|x}\}_{a=\pm 1}$ be projectors for each $x$, i.e., 
\begin{equation}
E_{a|x}=\frac{1}{2}(\openone+a\cdot\vec{x}\cdot\vec{\sigma}),
\end{equation}
with $a\cdot\vec{x}$ being the vector corresponding to projector $E_{a|x}$ in the Bloch sphere and $\vec{\sigma}=(\hat{X},\hat{Y},\hat{Z})$ being the Pauli matrices. Besides, the post-measurement states for each measurement event $a|x$ will be $E_{a|x}$. The temporal assemblage would be
\begin{widetext}
\begin{equation}
\begin{aligned}
\rho_{a|x}(t) &= p(a|x)~\mathcal{E}\left(E_{a|x}~\frac{\openone}{2}~E_{a|x}\right)\\
&=\frac{1}{2}\mathcal{E}\left(\frac{\openone}{2}\right)
+\frac{a}{8}\sum_{j=1}^3\text{tr}\left[\sigma_j[\mathcal{E}\left(E_{1|x}\right)-\mathcal{E}\left(E_{-1|x}\right)]\right]\sigma_j\\
&=\frac{1}{2}\cdot\frac{1}{2}\sum_{k=0}^3\text{tr}\left[\sigma_k\cdot\mathcal{E}\left(\frac{\openone}{2}\right)\right]\sigma_k
+\frac{a}{8}\sum_{j=1}^3\text{tr}\left[\sigma_j[\mathcal{E}\left(E_{1|x}\right)-\mathcal{E}\left(E_{-1|x}\right)]\right]\sigma_j\\
&=\frac{\openone}{4}
+\frac{1}{4}\sum_{k=1}^3\text{tr}\left[\sigma_k\cdot\mathcal{E}\left(\frac{\openone}{2}\right)\right]\sigma_k
+\frac{a}{8}\sum_{j=1}^3\text{tr}\left[\sigma_j[\mathcal{E}\left(E_{1|x}\right)-\mathcal{E}\left(E_{-1|x}\right)]\right]\sigma_j.
\end{aligned}
\label{Eq_temporal_assemblage}
\end{equation}
\end{widetext}

Then, by the definition of the pseudo density matrix $R$ in the main text, we can write down the pseudo density matrix in the Pauli bases:
\begin{widetext}
\begin{equation}
R=\frac{1}{4}
\left[\openone\otimes\openone
+\sum_{k=1}^3\text{tr}\left[\sigma_k\mathcal{E}\left(\frac{\openone}{2}\right)\right]\openone\otimes\sigma_k
+\frac{1}{2}\sum_{i,j=1}^{3}\text{tr}\bigg[\sigma_j[\mathcal{E}(E_{1|i})-\mathcal{E}(E_{-1|i})]\bigg]\sigma_i\otimes\sigma_j
\right].
\end{equation}
\end{widetext}
Finally, the target quantity $\text{tr}_\text{A}(E_{a|x}\otimes\openone~R)$ in Eq.~(\ref{App_Eq_TS_from_PDM}) would be
\newpage
\begin{widetext}
\begin{equation}
\begin{aligned}
\tr_\text{A}(E_{a|x}\otimes\openone~R) =&
\frac{1}{4}\tr_\text{A}(E_{a|x}\otimes\openone)
+\frac{1}{4}\sum_{k=1}^3\tr\left[\sigma_k\mathcal{E}\left(\frac{\openone}{2}\right)\right]\tr_\text{A}(E_{a|x}\otimes\sigma_k)\\
&+\frac{1}{8}\sum_{ij=1}^3\tr\bigg[\sigma_j[\mathcal{E}(E_{1|i})-\mathcal{E}(E_{-1|i})]\bigg]
\tr_\text{A}(E_{a|x}\otimes\openone\cdot\sigma_i\otimes\sigma_j)\\
=&\frac{\openone}{4}+\frac{1}{4}\sum_{k=1}^3\tr\left[\sigma_k\mathcal{E}\left(\frac{\openone}{2}\right)\right]\sigma_k
+\frac{1}{8}\sum_{ij=1}^3\tr\bigg[\sigma_j[\mathcal{E}(E_{1|i})-\mathcal{E}(E_{-1|i})]\bigg]\tr(E_{a|x}\sigma_i)~\sigma_j.
\end{aligned}
\end{equation}
\end{widetext}
Using the fact that $\tr(E_{a|x}\sigma_i)=a\cdot\delta_{x,i}$, the above equation will be the same as Eq.~(\ref{Eq_temporal_assemblage}). Since now we have the temporal assemblages, obtained from the pseudo density matrix, it is straightforward to obtain a set of temporal correlations $p(a,b|x,y)$.

We should note that from Eq.~(\ref{App_Eq_TS_from_PDM}), the way one obtains the temporal assemblage by performing measurement on the pseudo density matrix is merely a mathematical relation between $\rho_{a|x}(t)$, $E_{a|x}$, and $R$, instead of a physical system being \emph{measured}. This is different from the case in the standard spatial scenario that one obtains an assemblage by performing a set of local measurements on a subsystem of a quantum state. On the other hand, as we mentioned before, the reason to use the maximally mixed state as initial state is to obey the condition of no-signalling in time (NSIT).

\section{Proof of Proposition}{\label{App_Pro1}
To support the proposition, in the following we will show the partial transpose of pseudo density matrix $R$ is always positive semidefinite, i.e. $R^{T_{\text{A}}}\geq 0$, by considering the three standard quantum channels - the amplitude-damping channel, the phase-damping channel, and the depolarizing channel - which are often used to describe the dynamics of a system. Then, using the positive-partial-transpose (PPT) criterion~\cite{Horodecki96} and~\cite{Peres96}, it is easy to show that $R$ is separable.

The dynamics of a qubit undergoing the amplitude-damping, the phase-damping, and the depolarizing channels, can be respectively described by the following three Lindblad-form master equations:
\begin{subequations}
\begin{align}
\dot{\rho_\text{s}} &= \frac{\gamma_\text{A}}{2}\left( 2\sigma_-\rho_\text{s}(t)\sigma_+ - \sigma_+\sigma_-\rho_\text{s}(t) - \rho_\text{s}(t)\sigma_+\sigma_- \right),  \\
\dot{\rho_\text{s}} &= \frac{\gamma_\text{P}}{4}\left( 2\sigma_3\rho_\text{s}(t)\sigma_3 - \sigma_3^2\rho_\text{s}(t) - \rho_\text{s}(t)\sigma_3^2 \right),  \\
\dot{\rho_\text{s}} &= \frac{\gamma_\text{D}}{8}\sum_i\left( 2\sigma_i\rho_\text{s}(t)\sigma_i - \sigma_i^2\rho_\text{s}(t) - \rho_\text{s}(t)\sigma_i^2 \right)\label{Eq_master_D},
\end{align}
\end{subequations}
where $\rho_\text{s}$ is the standard density matrix of the qubit, $\{\gamma_i\}_{i=\text{A},\text{P},\text{D}}$ denote the decay rates of the dynamics in the different channels, and $\sigma_{+}$ ($\sigma_{-}$) is the creation (annihilation) operator. Assisted by the definition of the pseudo density matrix, one can obtain the pseudo density matrix in each scenario:
\begin{widetext}
\begin{subequations}
\begin{align}
R_\text{A} &=
\begin{pmatrix}
\frac{1}{2} e^{-t\gamma_\text{A}} & 0 & 0 & 0 \\
0 & \frac{1}{2}(1-e^{-t\gamma_\text{A}}) & \frac{1}{2} e^{-t\gamma_\text{A}/2} & 0 \\
0 & \frac{1}{2} e^{-t\gamma_\text{A}/2} & 0 & 0 \\
0 & 0 & 0 & \frac{1}{2}
\end{pmatrix},  \\
R_\text{P} &=
\begin{pmatrix}
\frac{1}{2} & 0 & 0 & 0 \\
0 & 0 & \frac{1}{2} e^{-t\gamma_\text{P}} & 0 \\
0 & \frac{1}{2} e^{-t\gamma_\text{P}} & 0 & 0 \\
0 & 0 & 0 & \frac{1}{2}
\end{pmatrix},  \\
R_\text{D} &=
\begin{pmatrix}
\frac{1}{4} (1+e^{-t\gamma_\text{D}}) & 0 & 0 & 0 \\
0 & \frac{1}{4}(1-e^{-t\gamma_\text{D}}) & \frac{1}{2} e^{-t\gamma_\text{D}} & 0 \\
0 & \frac{1}{2} e^{-t\gamma_\text{D}} & \frac{1}{4}(1-e^{-t\gamma_\text{D}}) & 0 \\
0 & 0 & 0 & \frac{1}{4}(1+e^{-t\gamma_\text{D}})
\end{pmatrix}.
\end{align}
\end{subequations}
\end{widetext}
It can be shown that the partial transpose of each above pseudo density matrix is always positive semidefinite, i.e., $R^{T_\text{A}}\geq 0$ for $t\in~(0,\infty)$. The fact that $f=0$ implies $R\geq 0$, indicating $R$ can be treated as a valid density matrix describing a qubit-qubit system. By using the positive-partial-transpose (PPT) criterion~\cite{Horodecki96} and~\cite{Peres96}: a density matrix $\varrho_\text{AB}$ describing a qubit-qubit (or a qubit-qutrit) system is separable if and only if its partial transpose $\varrho_\text{AB}^{T_\text{A}}$ is positive-semidefinite. In summary, we prove the proposition by the following steps:
\begin{equation}
R\geq 0 ~~\wedge~~ R^{T_\text{A}}\geq 0 ~~\wedge~~ \text{PPT criterion}~~~\Rightarrow~~~R~~~\text{is separable}.
\end{equation}

\section{Proof of Eq.~10 in the main text}\label{App_TSR_traceD}
Following the property of temporal steerable weight (TSW)~\cite{SLChen16_2}, one realizes that
\begin{equation}
\sigma _{a|x}^{T}-\sum_{\lambda }D_{\lambda }(a|x)\sigma _{\lambda }\geq 0,
\end{equation}%
where $D_{\lambda }(a|x)$ are the extremal deterministic values, $\lambda $
represents a local hidden variable, $x$ is the measurement basis, and $a$ is
the measurement outcome. Since $\sum_{a}D_{\lambda }(a|x)=1$, one has the
following
\begin{equation}
\sum_{a}\sigma _{a|x}^{T}-\sum_{\lambda }\sigma _{\lambda }\geq 0,
\end{equation}%
If we are limited to two measurement inputs and preparing the
assemblages with a classical way (without the off-diagonal terms), the
summation of the temporal assemblages $\sigma _{a|x}^{T}$ can be written as%
\begin{equation}
\sum_{a}\sigma _{a|1}^{T}=\left(
\begin{array}{cc}
\alpha  & 0 \\
0 & 1-\alpha
\end{array}%
\right) ,
\end{equation}%
\begin{equation}
\sum_{a}\sigma _{a|2}^{T}=\left(
\begin{array}{cc}
\beta  & 0 \\
0 & 1-\beta
\end{array}%
\right) .
\end{equation}%
Let us assume $\alpha >\beta $. The summation of the local hidden assemblage
$\overset{\sim }{\sigma _{\lambda }}=\sum_{\lambda }\sigma _{\lambda }$ that
can best mimic the temporal assemblages and fulfill the requirement of Eq.~(B2) is thus written as%
\begin{equation}
\overset{\sim }{\sigma _{\lambda }}=\left(
\begin{array}{cc}
\beta  & 0 \\
0 & 1-\alpha
\end{array}%
\right) .
\label{Eq_App_opt_TSW}
\end{equation}%
To prove that Eq.~\eqref{Eq_App_opt_TSW} is the optimal solution, one can add a non-negative number $\epsilon$ into the diagonal terms of the matrix in Eq.~\eqref{Eq_App_opt_TSW}. It is easy to see that $\text{Tr}(\tilde{\sigma}_\lambda)$ is maximum when $\epsilon=0$.
Therefore, the TSW is equal to the trace distance
between the two states $\sum_{a}\sigma _{a|1}^{T}$ and $\sum_{a}\sigma
_{a|2}^{T}$, i.e.
\begin{equation}
\text{TSW}=1-\text{Tr}(\overset{\sim }{\sigma _{\lambda }})=\alpha -\beta .
\end{equation}%
A similar argument can also be applied to the temporal steering robustness (TSR)~\cite{HYKu16}
with the following requirement
\begin{equation}
\sum_{\lambda }\sigma _{\lambda }-\sum_{a}\sigma _{a|x}^{T}\geq 0.
\end{equation}%
This leads one to write the summation of the local hidden assemblage as
\begin{equation}
\overset{\sim }{\sigma _{\lambda }}=\left(
\begin{array}{cc}
\alpha  & 0 \\
0 & 1-\beta
\end{array}%
\right) ,
\end{equation}%
and the corresponding TSR is written as
\begin{equation}
\text{TSR}=\text{Tr}(\overset{\sim }{\sigma _{\lambda }})-1=\alpha -\beta .
\end{equation}%
These conclude our proof that, in the classical scenario (no off-diagonal
elements), the temporal steering is equal to the trace distance between the summation of the elements of the temporal assemblage in different measurement settings.


\begin{thebibliography}{63}
\bibitem [{foo()}]{footnote}%
  \BibitemOpen
  \href@noop {} {\bibinfo  {journal} {These two Hilbert spaces are basically
  the same. However, due the the following discussion, e.g. take partial trace
  of $R$, it is necessary to use distinguishable subscripts.}\ }\BibitemShut
  {NoStop}%
\end{thebibliography}
%


\end{document}